\documentclass[final]{aipproc}

\layoutstyle{6x9}

\usepackage{epsfig}
\usepackage{amsmath}
\usepackage{bm}

\def    \ie             {{\em i.e.\/} }
\def    \etal             {{\em et al.\/}}
\def    \eg             {{\em e.g.}}
\newcommand{\ud}     {\mathrm{d}}

\newcommand{\gevsq}  {\;\mathrm{GeV}^2}
\renewcommand{\Im}{\mathop{\mathrm{Im}}}

\newcommand{\ceps}{\varepsilon}

\newcommand{\eq}[1]{Eq.(\ref{#1})}
\newcommand{\Eqs}[2]{Eqs.(\ref{#1}) and (\ref{#2})}

\newcommand{\as}{\alpha_{\scriptscriptstyle S}}
\newcommand{\sgls}{S_{\mathrm{GLS}}}
\newcommand{\sadler}{S_{\mathrm{A}}}

\begin{document}

\title{Nuclear Effects in Neutrino Structure Functions
}


\author{S. A. Kulagin}{
  address={Institute for Nuclear Research, 117312 Moscow, Russia}
}
\author{R. Petti}{
  address={Department of Physics and Astronomy, University of South Carolina, Columbia SC 29208, USA}
}

\begin{abstract}
We discuss calculation of nuclear corrections to the structure functions
for the deep-inelastic scattering of muon and (anti)neutrino. 
Our approach includes a QCD description of the
nucleon structure functions as well as the treatment of 
Fermi motion and nuclear binding, 
off-shell correction to bound nucleon structure functions,
nuclear pion excess and nuclear shadowing.
We emphasize the dependence of nuclear effects on the type and $C$-parity of
(anti)neutrino structure functions.
We also examine the interplay between different nuclear effects in the Adler and the
Gross--Llewellyn-Smith sum rules for nuclei.
\end{abstract}

\maketitle



\subsection{Introduction}

The study of nuclear effects in deep-inelastic structure functions (SF)
has by now a long and rich history.
The measurement of the ratio $\mathcal{R}_2(A/B)=F_2^A/F_2^B$ of SF of
two nuclei (usually a complex nucleus to deuterium)
in DIS experiments with charged leptons (CL) show that 
significant nuclear effects are present even in the scaling regime (``EMC effect'')
\cite{emc-effect-data}.
Many theoretical and experimental studies of nuclear effects on
structure functions followed, and an extensive
phenomenology has been developed, even though many questions about
the origin of the effect still remain.
Most of the direct experimental information on nuclear effects in the structure functions
come from the CL scattering data.
Future measurements with neutrino ($\nu$) and antineutrino ($\bar{\nu}$) beams 
are particularly interesting and would provide complementary information to CL data
\cite{Mangano:2001mj,minerva}.

The quantitative understanding of nuclear effects in neutrino scattering
is of primary importance for a correct interpretation  of experimental results
and for the evaluation of the corresponding uncertainties.
An accurate account of nuclear effects is
important in the determination of the parton distributions 
and the high-twist terms \cite{Alekhin:2003qq,Alekhin:2007fh},
the electroweak parameters in neutrino scattering experiments \cite{Kulagin:2003wz}, 
and also for the understanding of neutrino masses
and mixing. The next generation experiments would imply precision
measurements to disentangle small oscillation signals from neutrino and antineutrino
interactions on nuclei. This would require to improve our knowledge
of $\nu(\bar{\nu})$-nucleus cross sections in order to reduce systematic uncertainties. 

In this communication we summarize the results of recent studies of nuclear DIS
with charged leptons \cite{KP04} and neutrino \cite{KP07},
which were aimed to develop a consistent and quantitative
model of nuclear DIS applicable in the analyses of existing data and in
the interpretation of future experiments.

\vspace{-2ex}
\subsection{Nuclear structure functions in charged-lepton DIS} 

The approach of Ref.\cite{KP04} involves the
treatment of a number of mechanisms
in the regions of large and small values of the Bjorken variable $x$.
The physics scale of the separation between "small" and "large" $x$ regions
comes from the comparison of a characteristic longitudinal DIS correlation length 
$1/(Mx)$ in the laboratory system (see, \eg, \cite{IKL84,rev1}) 
with an average distance between bound nucleons. 
The DIS time is small at large $x$ ($x>0.1$), that
justifies the approximation of nuclear DIS  
in terms of incoherent scattering off bound protons and neutrons 
(the impulse approximation, or IA).
Important nuclear corrections in
this region are due to nuclear binding and nucleon momentum distribution (Fermi
motion).
At small $x$ the effects beyond the IA should be taken into account,
such as corrections due to scattering on nuclear pions (meson exchange currents) and
coherent multiple interactions of intermediate hadron (or quark-gluon) states
with bound nucleons that cause nuclear (anti)shadowing effect.
Summarizing, for the nuclear structure function
(to be specific we discuss $F_2$) we have different contributions:
\begin{equation}
\label{FA}
F_2^A = F_2^{\rm (IA)} + \delta_\pi F_2^{A} + \delta_{\rm coh} F_2^A,
\end{equation}
where the first
term in the right-hand side stand for the impulse approximation,
and $\delta_\pi F_2$ and $\delta_{\rm coh} F_2$ are the corrections
due to scattering off nuclear pion (meson) field and
coherent interaction of the intermediate virtual boson with nuclear target, respectively.

The nuclear SF in impulse approximation can be written as the sum of the proton ($\tau=p$) and neutron ($\tau=n$)
contributions
\begin{equation}
F_2^{\rm (IA)}(x,Q^2)
= \sum_{\tau=p,n}
\int [\ud k] \big(1+{k_z}/{M}\big) 
\mathcal{P}^\tau(\ceps,\bm{k})
F_2^\tau (x',Q^2,k^2)
\label{IA}
\end{equation}
where the integration is taken over the four-momentum of the bound nucleon
$k=(M+\ceps,\,\bm{k})$ and $[\ud k]=\ud\ceps\,\ud^3\bm{k}/(2\pi)^4$.
The energy and momentum distribution of bound nucleons is
described by proton (neutron) nuclear spectral function
$\mathcal{P}^{\smash{p(n)}}(\varepsilon,\bm{p})$, which is normalized to the
proton (neutron) number in a nucleus.
In the integrand $F_i^{\smash{p(n)}}$ are the structure
functions of bound proton (neutron), which
depends on the Bjorken variable $x'=x/(1+(\ceps+k_z)/M)$, momentum transfer
square $Q^2$ and also on the nucleon invariant mass squared
$k^2=(M+\ceps)^2-\bm{k}^2$.
We also note that \eq{IA} is written for the kinematics of the
Bjorken limit assuming that the momentum transfer is antiparallel to $z$ axis.
For the derivation and more general
expressions valid at finite $Q^2$ 
beyond the kinematics of the Bjorken limit see \cite{KP04}.


Although the calculation of Fermi motion and nuclear
binding corrections with realistic nuclear spectral functions 
provide a general trend of observed behavior of the
ratios $\mathcal{R}_2=F_2^A/F_2^D$ (EMC ratio) at large $x$ \cite{KP04}, 
the quantitative description of data is missing in IA. 
This indicates that the modification of the nucleon structure functions in
nuclear environment is an important effect.
In Ref.\cite{KP04} this effect is related to off-shell dependence,
\ie  the $k^2$ dependence, of structure functions in \eq{IA}.
Since characteristic energies and momenta of bound nucleons are small
compared to the nucleon mass, the off-shell effect can be treated as a linear
correction in the parameter $v=(k^2-M^2)/M^2$ 
to the structure function of the on-shell nucleon:
\begin{equation}
F_2(x,Q^2,k^2)=F_2(x,Q^2)[1+v\,\delta f_2(x,Q^2)].
\label{os}
\end{equation}
The function $\delta f_2$ describes the relative off-shell
effect. In Ref.\cite{KP04} this function was phenomenologically extracted from
analysis of data on the ratios $\mathcal{R}_2$.

Although the IA with off-shell correction
gives the leading contribution to nuclear SF, it is by no means complete.
It is known that the 
balance equation for the nuclear light-cone momentum is violated in the impulse approximation:
because of nuclear binding the nucleons do not
carry all of the light-cone momentum of the nucleus.
This indicates the presence of non-nucleon degrees of freedom in nuclei
which carry the missing light-cone momentum.
One possible solution is to consider the scattering off the virtual nuclear pions
that results in a correction to the structure function, $\delta_\pi F_2$ 
(for a review and more references see \cite{rev1,rev2}).
In effective $\pi$N theory with pseudoscalar coupling 
it can be shown that the nuclear pion distribution
indeed reconciles with the balance of the total nuclear light-cone momentum \cite{Ku89}.
In Ref.\cite{KP04} the nuclear pion correction was
calculated in terms of the convolution of the light-cone distribution
of nuclear pion excess in a nucleus and is the pion structure function
leading to a positive correction at small values of $x$
(about 5\% correction at the maximum at $x \sim 0.1$ for heavy nuclei).

At smaller $x$ ($x\ll 0.1$) the coherent effects in nuclear DIS are relevant
(see, \eg, \cite{rev1}). These effects are associated with the propagation 
of intermediate hadronic states in a nucleus. 
In particular, the interference of multiple scattering terms and 
the energy dependence of scattering amplitudes can cause both
the negative (shadowing) and positive (antishadowing) corrections 
depending on the region of $x$.
The rate of this effect depends on the scattering amplitude 
of the virtual hadronic states off the nucleon in forward direction.
In Ref.\cite{KP04} the interaction of virtual hadronic states with the
nucleon was described in terms of phenomenological scattering amplitude $a_{\rm eff}$.
The corresponding nuclear amplitude and correction to the structure function $\delta_{\rm coh}F_2$ was calculated
using the Glauber--Gribov multiple scattering theory.

The outlined approach was applied in the analysis of data
on the ratios $\mathcal{R}_2(A/B)=F_2^A/F_2^B$ of structure functions
of two nuclei \cite{KP04}. 
We perform a fit to experimental data by minimizing 
$\chi^2=\sum (\mathcal{R}_2^{\rm exp}-\mathcal{R}_2^{\rm th})^2/\sigma^2(\mathcal{R}_2^{\rm exp})$,
where $\sigma^2$ represents the uncertainty of experimental data and the sum includes all the data points.
The analysis includes data on $\mathcal{R}_2(A/B)$ for a variety of targets from D to Pb 
with $Q^2\ge 1\gevsq$ and in a wide region of $x$ (overall about 560 data points).
In this way we phenomenologically constrain the off-shell function $\delta f_2$ and
the effective amplitude $a_{\rm eff}$ from data.
To this end we used a 3-parameter model to parametrize the $x$ dependence of the off-shell function $\delta f_2$
and one phenomenological parameter to describe the $Q^2$ dependence of effective
cross section $\sigma_{\rm eff}=2\Im a_{\rm eff}$.
This approach leads to a very good agreement with data,
reproducing the observed $x$, $Q^2$, and $A$ dependencies 
of nuclear effects in the structure functions, with overall $\chi^2{\rm /d.o.f.}=459/556$.
In order to test the model we performed a number of fits to different sub-sets of nuclei in the region
from $^4$He to ${}^{208}$Pb. The results are compatible within the
uncertainties with the result of the global fit thus indicating an
excellent consistency between the model and the data for all nuclei. 

\vspace{-3ex}
\subsection{Nuclear corrections in neutrino DIS}

In contrast to the CL case, the neutrino interactions with nuclei involve both
the vector and the axial current and charge exchange processes. For this reason
the results on CL structure functions can not be automatically taken over the neutrino case. 
Nevertheless, there is certain similarity in nuclear scattering mechanisms
for high-energy CL and neutrino interactions.
In particular, \eq{IA} can be applied to calculate neutrino-nuclear SF in impulse
approximation. 
In the analysis of neutrino and antineutrino SF for heavy nuclei
it is useful to explicitly separate the isoscalar and isovector
parts in \eq{IA} by introducing the isoscalar, 
$\mathcal{P}^p+\mathcal{P}^n=A \mathcal{P}_0$,
and the isovector, 
$\mathcal{P}^p-\mathcal{P}^n=(Z-N) \mathcal{P}_1$, nuclear spectral functions.
The $Z$ and $N$ are the proton and neutron number and the reduced spectral functions $\mathcal{P}_{0}$ and
$\mathcal{P}_{1}$ are normalized to unity.
Then we can write \eq{IA} as
\begin{equation}
\label{IA:01}
F_i^A/A = \frac12 \left\langle F_i^{p+n} \right\rangle_0 +
	\frac{\beta}{2} \left\langle F_i^{p-n} \right\rangle_1,
\end{equation}
where $i$ labels the SF type, $F_i^{p\pm n}=F_i^p \pm F_i^n$ and
the parameter $\beta=(Z-N)/A$ describes the excess of protons over neutrons in a nucleus.
In \eq{IA:01} we use the contracted notations of the integration in \eq{IA} with reduced spectral
functions $\mathcal{P}_0$ and $\mathcal{P}_1$, respectively.
In the kinematics corresponding to the Bjorken limit the integrands in
\Eqs{IA}{IA:01} are similar for different types of SF ($i=T,L,2,3$).
However, the finite $Q^2$ effects in nuclear convolution depend on the SF type.
In addition, the effect of the nucleon transverse momentum leads to the mixing of
$F_T$ and $F_L$ contributions to order $\bm{p}_\perp^2/Q^2$ (for more detail see \cite{KP07}).


Equation (\ref{IA:01}) has a convenient form for the calculation of neutrino SF with definite $C$-parity.
In particular, we apply it to $\nu \pm \bar\nu$ combinations.
For the $C$-even combination we have 
$F_i^{\smash{ (\nu+\bar\nu)p}}=F_i^{\smash{(\nu+\bar\nu)n}}$, because of the isospin
symmetry. For this reason the isovector term in \eq{IA:01} should vanish.
However, one should remark that the isospin relations for the structure
functions for neutrino CC scattering are violated by the mixing of different
quark generations and the $c$-quark mass effect, even in the presence of
exact isospin symmetry for the parton distributions. This effect results in
the nonzero difference $F_i^{\smash{(\nu+\bar\nu)(p-n)}} \propto \sin^2\theta_C$ with
$\theta_C$  the Cabbibo mixing angle. 
Nevertheless, in heavy nuclei the contribution of isovector term is accompanied by a small parameter $\beta$ 
and, therefore, it is a good approximation to keep only the isoscalar term in \eq{IA:01}.

The application of \eq{IA:01} to the $\nu{-}\bar\nu$ asymmetry of $F_2$ and $xF_3$ requires somewhat
more attention. We first consider $F_2^{\nu-\bar\nu}$ (similar discussion also
applies to $F_T$ and $F_L$). This structure function is $C$-odd and
dominated by the isovector quark distributions.
The isospin symmetry suggests
$F_2^{\smash{(\nu-\bar\nu)p}}=-F_2^{\smash{(\nu-\bar\nu)n}}$. Therefore,
the first term in the right side of \eq{IA:01} vanishes
and the nuclear structure function is determined by the second (isovector) term. 
However, as mentioned above, the SF isospin relations are violated 
and $F_2^{\smash{(\nu{-}\bar\nu)(p+n)}}\propto\sin^2\theta_C$.
Surprisingly, this effect should not be
neglected in the analysis of nuclear corrections for the difference $\nu-\bar \nu$
even in the first approximation. This is because the isovector
contribution in \eq{IA:01} is suppressed by the factor $\beta$ and
thus the relative contribution of $F_2^{\smash{(\nu{-}\bar\nu)(p+n)}}$ is enhanced \cite{KP07}.
The difference $xF_3^{\nu-\bar\nu}$ is $C$-even
and includes the (isovector) contribution from the light quarks
and the $s$-quark contribution to the isoscalar part. 
For the isoscalar nucleon the difference 
$xF_3^{\smash{(\nu{-}\bar\nu)(p+n)}}$ is driven by the $s$-quark distribution 
[in the leading approximation in the parton model $2x(s+\bar s)$]. Because
$\beta$ is a small parameter in heavy nuclei,
the contribution from the isovector term in \eq{IA:01} is suppressed and,
therefore, $xF_3^{\nu-\bar\nu}$ in nuclei is dominated by the strange quark
contribution.

Similar to the CL case,
the IA should be corrected for the nuclear pion effect (PI) and the effects associated 
with multiple coherent interactions of intermediate hadronic states in nuclear environment (NS).
The PI correction can be treated similar to the CL case.
In the convolution approximation we find that the PI corrections
to neutrino and CL $F_2$ are similar, and the corresponding correction
to $xF_3$ can be neglected. The coherent nuclear correction $\delta_{\rm coh}F_i$
essentially depends on the type of the structure function $i$ and
on the $C$-parity \cite{KP07}.
In general, the NS corrections $\delta_{\rm coh}F_{2,3}^{\nu\pm\bar\nu}$ can be described in terms of four
effective amplitudes $a^{\smash{(I,C)}}_{\rm eff}$, where $I=0,1$ corresponds to
the isoscalar (isovector) nucleon configuration and $C=\pm 1$ is $C$-parity.
For comparison, the corresponding CL correction is determined by
a single $a^{\smash{(0,+)}}_{\rm eff}$ amplitude.  
\vspace{-2ex}

\begin{figure}[htb]
\hspace*{5mm}\epsfig{file=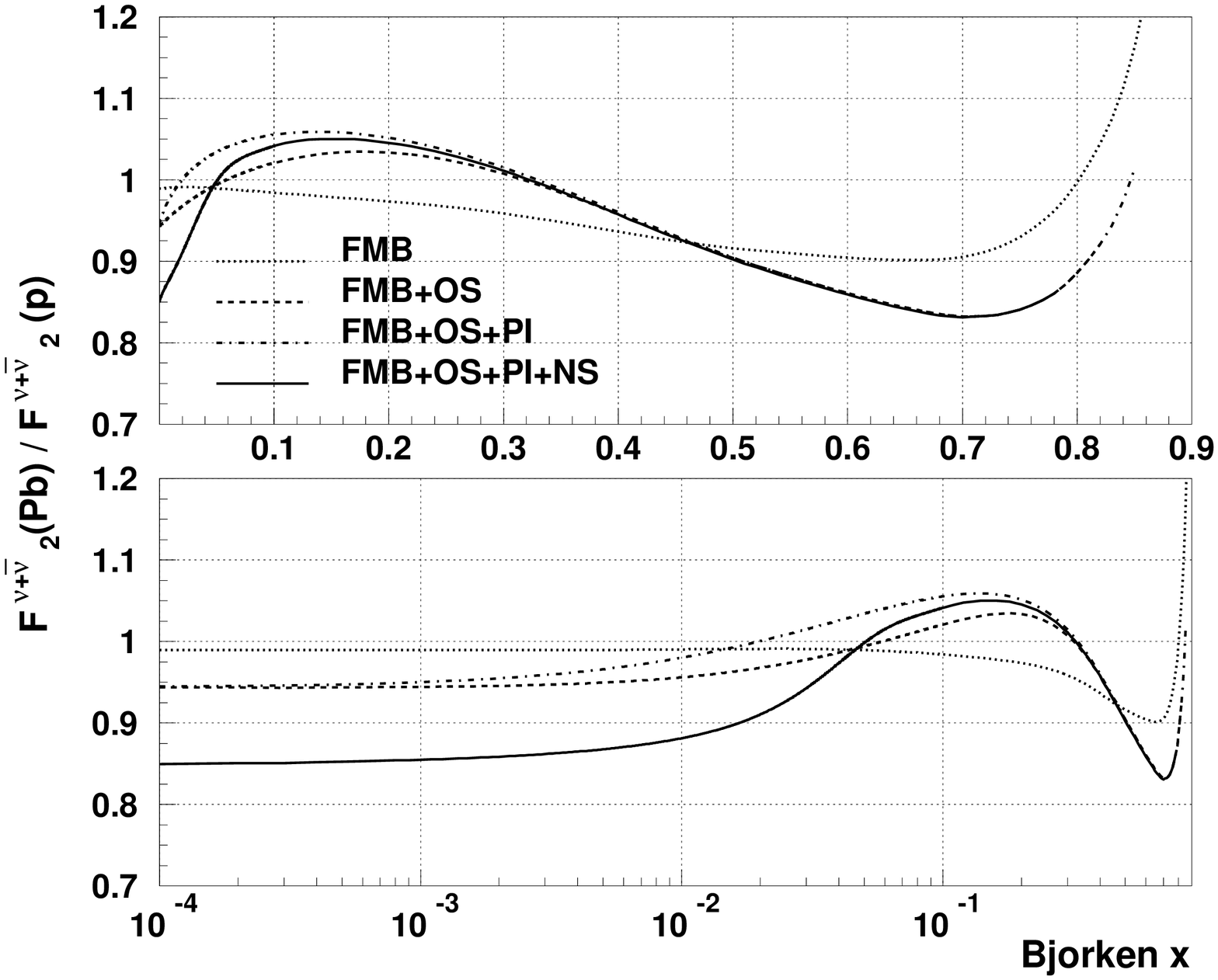,width=0.55\textwidth}%
\hspace*{-5mm}\epsfig{file=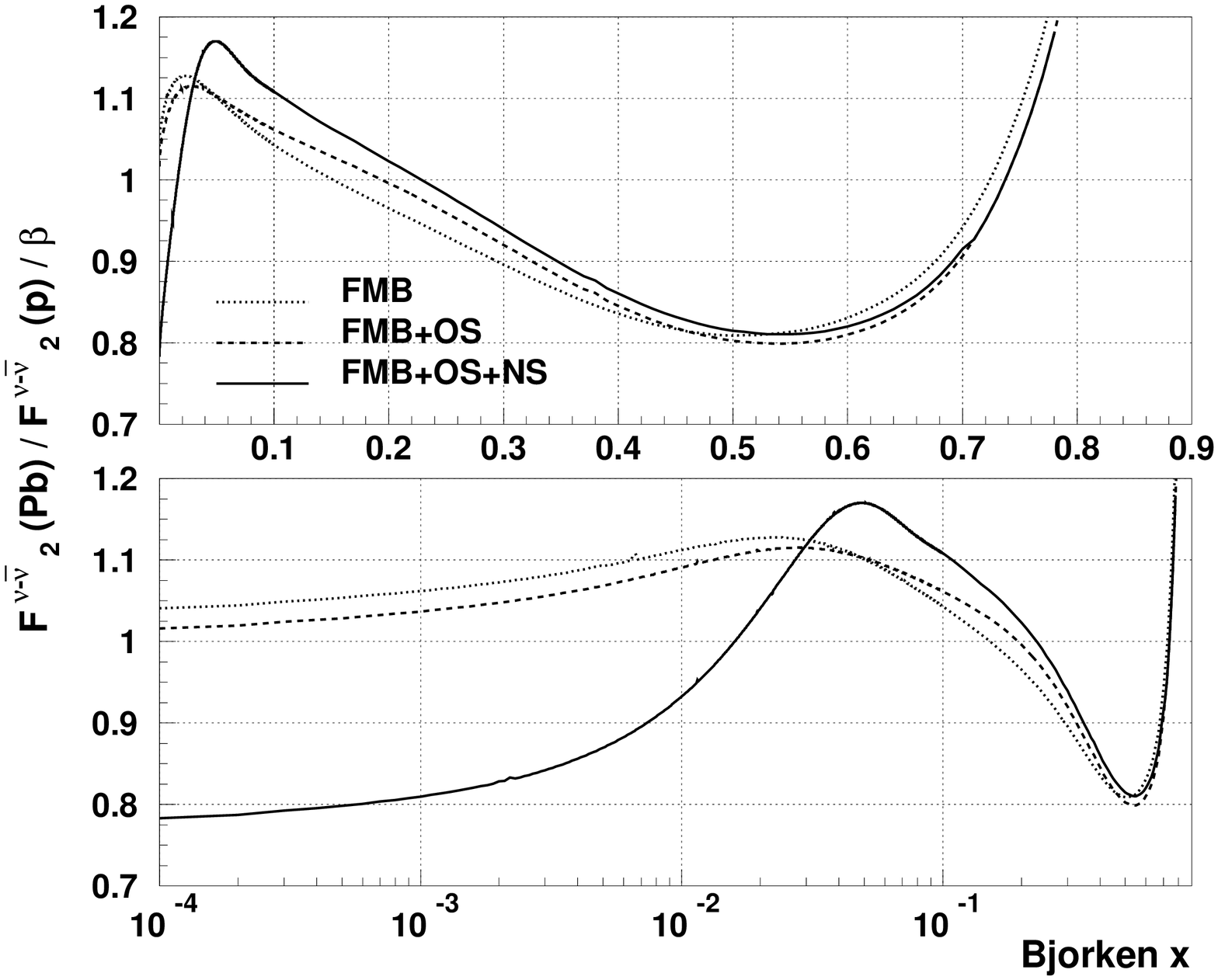,width=0.55\textwidth}
\caption{%
Different nuclear effects calculated for ${}^{207}$Pb at $Q^2=5\gevsq$
for $C$-even and $C$-odd combinations of neutrino SF (per one bound nucleon).
Shown are the ratio
$F_2^{\smash{(\nu+\bar{\nu})}A}/F_2^{\smash{(\nu+\bar{\nu})}p}$ (left panel)
and
$F_2^{\smash{(\nu-\bar{\nu})}A}/(\beta F_2^{\smash{(\nu-\bar{\nu})}p})$ (right panel).
The labels on the curves
correspond to effects due to Fermi motion and nuclear binding (FMB),
off-shell correction (OS), nuclear pion excess (PI) and coherent nuclear
processes (NS).
}
\label{fig:R}
\end{figure}

Figure~\ref{fig:R} illustrates different nuclear corrections to
the $C$-even and $C$-odd SF $F_2^{\nu \pm\bar{\nu}}$ of
${}^{207}$Pb calculated in our model \cite{KP07}.
The result for $F_2^{\nu+\bar{\nu}}$ resembles the EMC effect for $F_2^{\smash \mu}$ 
(for comparison, see Fig.3 of Ref.\cite{KP04}).
At large $x$ the nuclear correction is driven by FMB and OS effects. We
recall that the off-shell effect, which modifies the structure
functions of bound nucleon, is important to change the slope of the
EMC ratio bringing it close to the CL data.
At intermediate $x$ values we observe some
cancellation between the PI and the NS corrections. 
This latter is the dominant effect for $x<0.05$, providing a
suppression of the nuclear structure function.

The corresponding nuclear
corrections for the difference $F_2^{\nu - \bar{\nu}}$ are shown
after normalizing them to $\beta = (Z-N)/A$.
The shape and the magnitude of nuclear corrections are somewhat different
from the $C$-even case.
Note that the coherent nuclear effects 
are particularly pronounced in the $C$-odd channel,
leading to a larger shadowing effect at $x<0.005$.
The enhancement at $0.02<x<0.2$ (antishadowing) with the maximum at $x\sim 0.05$ 
is a combined effect due to nuclear
smearing (FMB) and a constructive interference in coherent nuclear
correction which is driven by the real part of the effective amplitude 
$a^{\smash{(1,-)}}_{\rm eff}$.

\vspace{-3ex}
\subsection{Neutrino DIS sum rules for nuclei}

The DIS sum rules reflect the underlying symmetry of interactions and
put rigid normalization constraints on the structure functions.
In the study of nuclear DIS the sum rules play a special role providing
a link between different nuclear effects.

In this Section we briefly discuss the Adler and the Gross--Llewellyn-Smith (GLS)
sum rules for nuclei (for more detail see \cite{KP07}). 
The Adler sum rule provides the relation between the integrated difference of the
isovector combination $F_2^{\nu - \bar\nu}$ to the isospin of the target \cite{AdlerSR}:
\begin{equation}
\label{ASR}
\sadler=\int_0^{M_A/M} 
\ud x\ F_2^{\bar\nu -\nu}(x,Q^2)/(2x)
= 2\,I_z,
\end{equation}
where the upper integration limit, the ratio of the nuclear and the
nucleon masses $M_A/M$, is the maximum possible value of the Bjorken variable
$x=Q^2/(2Mq_0)$ for the nuclear target and $I_z$ is the
projection of the target isospin vector on the quantization axis ($z$ axis).
In the quark parton model the Adler sum is the difference between the
number of valence $u$ and $d$ quarks of the target. The Adler sum rule
is valid beyond the parton model and survives the strong interaction effects
because of the underlying isospin symmetry of strong interaction.
However, in the derivation of the Adler sum rule the effects of
non-conservation of the axial current that limits its applicability to the region of
relatively large $Q^2$. 
Also Eq.(\ref{ASR}) should be corrected for
Cabibbo mixing effect (see, \eg, \cite{IKL84}).

For the proton $\sadler(p)=1$ and for the neutron $\sadler(n)=-1$.
For a generic nucleus of $Z$ protons and $N$ neutrons the Adler
sum rule reads (we consider the nuclear structure functions per one nucleon):
\begin{equation}
\label{ASR:A}
\sadler=(Z-N)/A=\beta.
\end{equation}
We now discuss in turn the contributions to $\sadler$ from different
nuclear effects.
First we calculate the Adler sum rule for nuclear SF in impulse approximation.
By performing the direct integration by $x$ of \eq{IA:01} 
and using the isospin relations for the proton and neutron structure
functions we verify that the FMB correction exactly cancels out
and we obtain (\ref{ASR:A}).
Also, the nuclear pion correction to $\sadler$ vanishes because of charge conservation. 
However, the OS and NS contributions to $\sadler$ 
are both individually finite. The requirement of cancellation between these corrections
provides important constraint on the isovector part of nuclear effects.
We note that the NS correction to $F_2^{\nu-\bar\nu}$ depends on
effective scattering amplitude $a^{\smash {(0,+)}}_{\rm eff}$ and 
$a^{\smash {(1,-)}}_{\rm eff}$. The amplitude $a^{\smash {(0,+)}}_{\rm eff}$
also determines the NS corrections to $F_2^\mu$ and was phenomenologically
determined in the analysis of \cite{KP04}. 
The  Adler sum rule helps to fix the Re/Im ratio of $a^{\smash {(1,-)}}_{\rm eff}$ \cite{KP07}.
Note also that the Adler and the GLS sum rules were used in \cite{KP07} in order
to evaluate the effective cross section $\sigma_{\rm eff}=2\Im a^{\smash {(0,+)}}_{\rm eff}$ 
governing the NS effect at high $Q^2$.

The GLS sum rule is the integrated structure function $F_3^{\nu+\bar\nu}$
\begin{equation}
\label{GLS}
\sgls = \int_0^{M_A/M} 
\ud x\, F_3^{\nu+\bar\nu}(x,Q^2)/2,
\end{equation}
where we write the GLS integral per one nucleon for a generic nuclear
target. In the quark parton model the GLS sum gives the number of
valence quarks of the target $\sgls=3$ \cite{GLS}.
Since the number of valence quarks is directly related to the baryon number,
in the parton model
$\sgls$ is independent of nuclear target.
However, in QCD the direct relation between the baryon current and $\sgls$
only holds in the leading twist approximation (LT) and in 
the leading order (LO) in strong coupling constant $\as$.
In contrast to the Adler sum rule,
$\sgls$ depends on $Q^2$ and is affected by QCD radiative
corrections, target mass and the higher-twist effects.
In general, $\sgls$ also depends on nuclear target.

\begin{figure}[ht]
\epsfig{file=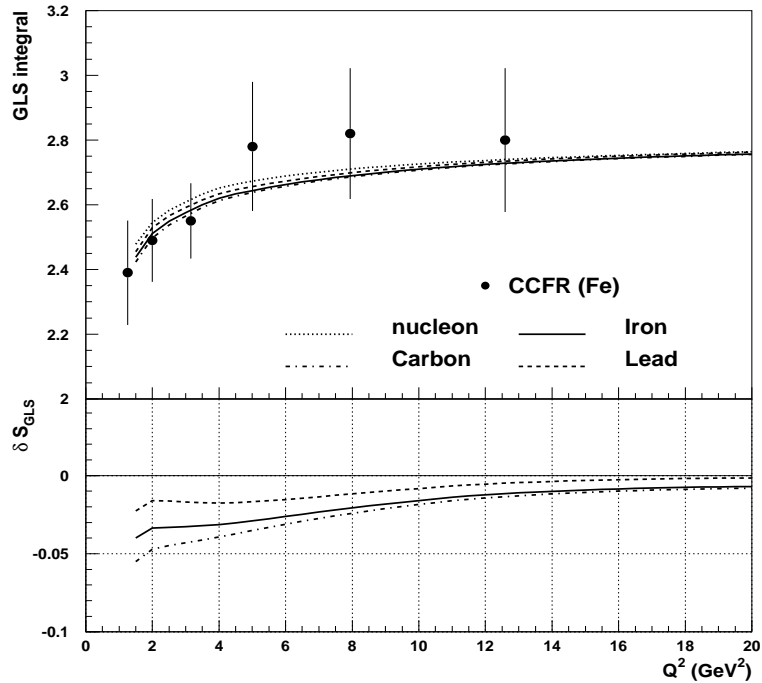
,width=0.75\textwidth
,height=0.7\textwidth
}
\caption{%
The GLS integral $\sgls$ (upper panel) 
and its variation with respect to the nucleon $\delta\sgls$ (lower panel)
calculated in \cite{KP07} for different nuclear targets as a function of $Q^2$.
Data points are CCFR extraction of the GLS integral for iron \cite{CCFR:GLS}.
}
\label{fig:gls}
\end{figure}

We explicitly separate nuclear corrections to
the GLS integral as $\delta \sgls=\sgls^{\smash A}- \sgls^{\smash N}$, where
$\sgls^{\smash N}$ refers to the GLS integral for the nucleon (proton)
and calculate the GLS integral (\ref{GLS}) for a number of nuclei as a function of $Q^2$
\cite{KP07}. The results are shown in Fig.~\ref{fig:gls}.

\vspace{-4ex}

A few comments are in order.
It can be verified analytically using \eq{IA} that $\delta\sgls=0$ in IA \cite{Ku98},
\ie the FMB correction to the GLS integral vanishes, similar to the Adler sum rule.
The cancellation between phenomenological OS and NS corrections at high $Q^2$ was verified in \cite{KP04}
(the normalization of nuclear valence quark distribution was one of the constraints of analysis \cite{KP04}).
However, at low $Q^2$ the cancellation between different nuclear effects in $\sgls$ is not exact.

The results of Fig.~\ref{fig:gls} were calculated with the LT nucleon SF
using the NNLO coefficient functions and the NNLO parton distributions of 
\cite{a02} and also include the target mass correction. 
As can be seen from Fig.~\ref{fig:gls} the correction $\delta\sgls$
is negative and of the order of 2\% at low $Q^2$ and decreases progressively with $Q^2$
and the results for different targets are similar.
This is mainly due to the explicit $Q^2$ dependence
of effective cross section describing the
nuclear shadowing effect.
We note that phenomenological cross section effectively
incorporates contributions from  all twists since it is extracted from
data.
Higher twists are known to be important at low $Q^2$ and
for this reason we should not expect the exact cancellation between
the OS and NS corrections in this region.
We also note that the general
$Q^2$ dependence for ${}^{56}$Fe is in agreement with the CCFR measurement \cite{CCFR:GLS}.

The nuclear effects somewhat modify the 
$Q^2$ evolution of the GLS integral. For this reason nuclear corrections should be
taken into account for precise extractions of $\as$
from the GLS sum rule.

\vspace{-3ex}
\subsection{Acknowledgments}
\noindent
S.K. thanks the Organizing Committee of the NuInt07 Workshop for 
warm hospitality and local support. 
A partial support from the Russian Foundation for Basic Research,
projects no. 06-02-16659 and 06-02-16353, is acknowledged.
R.P. thanks USC for supporting this research.

\end{document}